\documentclass[twocolumn,twoside]{IEEEtran} 

\usepackage{mymacros}
\usepackage{times,mathptm}
\usepackage[final]{graphicx}

\newcommand{\note}[1]{\newline \emph{Note: #1}}
\newcommand{\startadc}{\textsc{startadc}}
\newcommand{\savedata}{\textsc{savedata}}
\newcommand{\flushdata}{\textsc{flushdata}}
\newcommand{\abort}{abort}
\newcommand{\trigger}{trigger}

\begin{document}

\title{An Upgraded Data Acquisition System for the Balloon-Borne
       Liquid Xenon Gamma Ray Imaging Telescope LXeGRIT
}

\author{Elena~Aprile, Alessandro~Curioni, Karl-Ludwig~Giboni, Uwe~Oberlack,  
	\thanks{E.~Aprile (age@astro.columbia.edu), A.~Curioni, K.~L.~Giboni, 
                and \mbox{U.~Oberlack} are with 
                Columbia Astrophysics Laboratory, 
                Columbia University, New York, NY 10027 USA}%
        and Sandro~Ventura
	\thanks{S.~Ventura is with INFN / University of Padua, Italy.}
}

\markboth{IEEE Transactions On Nuclear Science --- Manuscript submitted November 26, 2000}
{E.Aprile et al.: An Upgraded Data Acquisition System for LXeGRIT}
\maketitle

\begin{abstract}
LXeGRIT is a balloon-borne Compton telescope for \MeV\ \g-ray astrophysics,
based on a liquid xenon time projection chamber (LXeTPC) with charge and light
read-out. The first balloon flights in 1997 revealed limitations of the trigger
electronics and the data acquisition (DAQ) system, leading to their upgrade. New
electronics was developed to handle the xenon scintillation light trigger. The
original processor module was replaced by a commercial VME processor. The
telemetry rate was doubled to $2 \times 500$~kbps and onboard data storage on
hard disks was implemented. Relying on a robust real-time operating system, the
new DAQ software adopts an object oriented design to implement the diverse tasks
of trigger handling, data selection, transmission, and storage, as well as DAQ
control and monitor functions. The new systems performed well during two flights
in Spring 1999 and Fall 2000. In the 2000 flight, the DAQ system was able to
handle 300 -- 350~triggers/s out of a total of about 650~Hz, including charged
particles.
\end{abstract}


\section{Introduction}

LXeGRIT is the first Liquid Xenon Time Projection Chamber used 
outside a laboratory, on a balloon borne platform 
\cite{EAprile:00:spie00:performance,EAprile:00:spie00:flight99}. The advantages 
of a large homogeneous detector as Compton Telescope for astrophysics 
justify the complex read-out system needed to acquire the complete 
spatial, temporal, and energy information of any ionizing event. 
Compared with other balloon borne scientific instruments, the 
TPC generates an enormous data rate, which after acquisition, 
has to be processed for background rejection, partial analysis 
as high level trigger, as well as packaging for either on-board 
storage or transfer via telemetry to ground. The front-end electronics, 
acquiring the data, was custom built, and so was the original 
read-out processor. The data acquisition system showed some severe 
shortcomings during the first engineering flights in 1997. The 
analog and digital front-end electronics was designed to fit 
the exact specifications and peculiarities of both the detector 
and the application. Most limitations were introduced by the 
custom built data processing system. Recognizing the advances 
and the availability of powerful computer systems, it was decided 
to replace the existing unit with a commercial device. Additionally, 
a new unit was introduced to handle the trigger signals, since 
the original circuit did not allow sufficient control over the 
trigger decisions, and also did not provide all the rates necessary 
to derive flux values.

The advantages of the new DAQ system are especially obvious during 
the development phase. The architecture of the data paths on 
the computer boards are optimized for efficient information transfer 
between the processor and the various communication interfaces. 
The computer architecture is backed by a powerful operating system. 
The system is also equipped with a Fast-Ethernet port, which provides a
high throughput link for control and data taking in
the laboratory. Two high speed RS-485 serial ports 
serve to transmit data on fast telemetry channels, whereas a 
SCSI interface allows to connect hard disks for onboard storage 
of large amounts of data. Most importantly, a VME port makes it 
possible to easily interface a variety of different data sources.

Although adequate for the present instrument, the processor system 
is used close to its capacity. Larger detectors, or even higher 
data rates would require multiple processor systems with more 
online computer power for online reconstruction of the events. 
The present system, besides providing valuable scientific data 
for gamma-ray astrophysics, also shows the way for the development 
of future more complex systems.

\section{System Hardware}

\subsection{Front-End Electronics and Flash-ADC (FADC) System}

The front-end and FADC system of the LXeGRIT instrument is described in
\cite{EAprile:98:electronics}. Here we recall its main features
before focusing on the system upgrades. The front-end electronics converts
the charge signals from the 124 induction wires (62 X- and 62 Y-wires) and
the 4 anodes into voltage pulses. Each channel has a charge sensitive
preamplifier, which drives the twisted pair line to the
digitizer system. The digitizers convert the analog signals
into a digital history of the ionizing event. The FADC system consists of 17
printed circuit boards housed in a standard VME-crate: 16 Gamma-Ray (X-Y)
Induction Signal Processor (GRISP) boards with 8 channels each to handle
the 124 wire signals, and one Gamma-Ray Anode Signal Processor (GRASP) board
to handle the 4 anode signals.

The X-Y wire signals are digitized with 8~bit precision at a rate of 5~MHz. The
information is stored in a dual port random access memory (DRAM). The depth of
this buffer is 256 samples, corresponding to 51.2~\us, which covers the maximum
drift time in the TPC of about 40~\us\ for a drift velocity of $\sim
2$~mm/\us. The charge collection signals from the 4 anode channels are digitized
at the same rate with 10~bit precision, for better energy determination with a
large dynamic range.

For each channel, the digital signal is passed through a comparator to
record the sample number when a software-set threshold is exceeded. The
recording of the threshold crossing point facilitates locating useful
information and can be used to reduce the data amount and to accelerate the
data read-out process.  Each GRISP board with at least one channel above
threshold issues a signal that sets a flag in a 16~bit register, which was
located on the microprocessor board in the original design, and is now
located on a separate board (``latch card'') within the crate.

The GRASP board can send 3 different interrupt requests to the processor:
\startadc\ and \savedata\ signal the start and the completion of
event digitizing, while \flushdata\ signals that the process was
interrupted by a second trigger, the system aborted the data recording,
and is ready to accept a new event. In the new design, these interrupts are
registered on the ``latch card'' mentioned above and read out by
the external processor. The GRASP board can also start an event
digitizing process on command from the read-out processor, independent of
an external trigger. These test triggers are used to determine baselines
and noise conditions on anodes and wires.

The front-end and FADC system of the LXeGRIT instrument has remained
unchanged from the original design, with the exception of the trigger
electronics. The circuitry amplifying and discriminating the signals from the 4
UV-sensitive photomultiplier tubes (PMTs), originally on the GRASP board, has
been replaced by new electronics. Event recording is
now triggered by a fast TTL pulse signaling the start of the event. The
recording is pre-triggered, but will be stopped if a second trigger pulse
signals the occurrence of a second event within 40~\us, while the charges
of the first event are still drifting in the sensitive volume of the
TPC. In this case, both events are rejected.

\subsection{Read-Out Processor}

Since the connections to the GRISP/GRASP boards followed the VME standard to a
large extent, it was natural to choose a VME processor board. The final choice
was a Motorola MVME 2700 coupled to a communication interface MVME761 Transition
Module. Not all connections in a standard VME bus are used by the GRISP/GRASP
FADC system, and some bus lines were assigned a different meaning. The
processor could therefore not be housed in the same crate. It is located instead
in a separate box, together with the communication board.

An interface board was developed to buffer the data and address 
lines, and also to emulate the correct timing of the handshake 
signals for data transfer. This is necessary to adapt the synchronous 
read and write cycles of the FADC memories to the inherent asynchronous 
operation of a standard VME bus. During data acquisition most 
of the operations on the bus are read cycles from the FADC memories. 
These cycles were therefore kept as short as possible (250~ns) 
to obtain the required data transmission rate.

Most of the data words to be read from the GRISP/GRASP boards are digitized
waveforms, which are stored in consecutive locations in memory. Block transfers
are thus a natural choice to increase data throughput compared to single
reads. Initial tests with block transfers, however, revealed that the processor
board does not keep the address lines stable during the full transfer, as this
is not required by the VME standard. The address lines had therefore to be
latched with each address strobe to achieve the higher transfer rate.

The FADC system interface board is connected to the VME port 
of the processor via the VME Junction. This circuit buffers the 
lines and allows for the connection of additional instruments 
to the VME bus. Presently there is one such instrument, the trigger 
logic system.

After processing the data, the processor can either send the events via two fast
serial ports to the science data transmitters, or via the SCSI port to two 36~GB
hard disks for storage onboard (two 9~GB disks in the 1999 flight). Although the
data can be stored on disk much faster than transmitted to ground via telemetry,
the data might be lost in case of a bad landing. To guarantee a good sample of
science data even in this case and to allow online control of the data
acquisition and thus the tuning of data taking parameters, a subset of the
acquired events is transferred to the two fast serial ports. After level
conversion, the data are sent by two transmitters to ground, with a throughput
of $2 \times 500$~kbits/s. The rate could be increased by about a factor of two,
but the analog tape drives of the National Scientific Balloon Facility (NSBF),
providing a backup copy of the science data, are not foreseen for such high
rates.

\begin{figure*}
\centering
\includegraphics[bb=105 344 396 749,angle=90,clip]{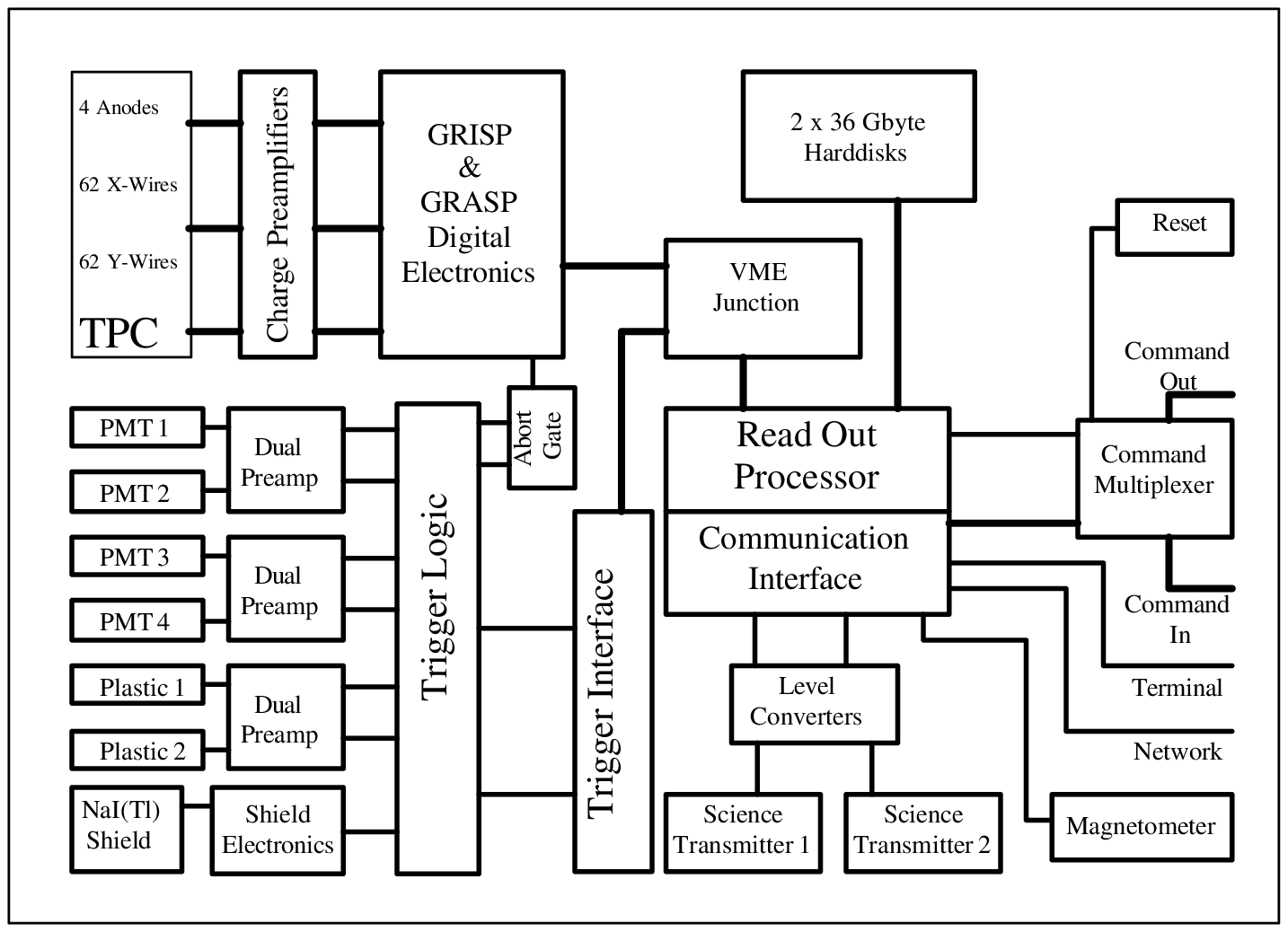}
\caption{\label{f:readout} Diagram of the interconnections of the LXeGRIT
read-out electronics.}
\end{figure*}

Other connections to the processors are: the magnetometer and tiltmeter
which provide directional information of the LXeGRIT instrument via 
a slow serial port; a terminal and an Ethernet connection used for
operation support while the payload is in the laboratory. During the flight,
the processor is controlled by 16~bit command words received via the
Consolidated Instrumentation Package (CIP), the standard NSBF package to
control instruments during balloon flights.  The commands are provided to
the parallel port of the processor via the command multiplexer, which
formats the 16~bit word into 2 bytes to be read consecutively. The
interconnections of the LXeGRIT read-out electronics are shown
schematically in Fig.~\ref{f:readout}.

\subsection{Trigger Logic}

As a pre-triggered digitizer system, the GRISP/GRASP boards requires a fast
signal to start recording an event. This signal is derived from the fast xenon
scintillation light detected by four
UV-sensitive PMTs, which view the sensitive volume of the TPC through quartz
windows. Originally, the system was triggered on a logical OR of the four PMT
signals, above a given threshold.  The trigger was then vetoed with the signal
from plastic counter(s) above and from the NaI(Tl) shield sections around and
below the LXeTPC. Timed gates removed double triggers. This system was
insufficient, mainly because no record was kept of the various signal rates,
only the four single PMT rates were stored together with other housekeeping
information. Since events can produce a signal on more than one PMT, and
triggers can be vetoed or rejected as double events, the information was not
sufficient for a rate calculation. A separate trigger electronics was therefore
custom built, providing the FADC system with a TTL pulse to start the read-out.

After amplification, the signals from the 4 PMTs are passed through 4 window
discriminators. The lower threshold of the window rejects noise pulses and can
also be used to introduce an energy bias by requiring a minimum energy deposited
in the TPC. The reason for the upper threshold, which is optional, was to
discriminate against charged particle tracks. A charged particle deposits about
3.9 \MeV/cm in liquid xenon, i.e., a total energy well in excess of the typical
gamma-ray energies of interest for observations of cosmic sources during the
flights. The OR of the PMT signals still generates the trigger for the FADC
system, unless it is vetoed or it is preceded by a PMT signal above the lower
discriminator threshold in the previous 40~\us\ (double or multiple events).  A
veto signal is generated by the OR of plastic and NaI(Tl) scintillators, used to
reject cosmic rays, and \g-rays entering the TPC from the side or from
below. For the 2000 flight, all shields were removed, thus no PMT signals were
vetoed.  In case of multiple events, an \abort\ signal is issued to stop the
FADC recording of the first trigger.  Since the FADC system does not have a
separate input for the \abort\ signal, event recording can only be stopped by a
second trigger-like signal. If, however, a first PMT-OR signal did not result in
a \trigger\ signal, either because it was vetoed or because the signal surpassed
the upper discriminator threshold, the occurence of a subsequent \abort\ signal
would in fact trigger the digitizer system.  An \abort\ gate was therefore
introduced to filter out all \abort\ signals not preceded by a \trigger.

A set of 16 counters, automatically reset once every second, register all
signals at various locations of the trigger logic, providing the means to
calculate the flux of events and the rejection rates. The rates from the
counters are also a very good monitor of the trigger system.  The trigger
electronics unit is connected to the VME bus of the read-out processor via an
interface. Thus, the processor can read the 16 counters, set the window
discriminator thresholds, and set the operation mode. Different operation modes
can be enabled: the veto signals can be switched on and off, the upper level of
the window discriminators can be turned off, and the veto can be replaced by a
coincidence, effectively triggering on charged particle tracks for debugging
purposes, or to study the spatial resolution of the detector in the laboratory.

\subsection{Mechanical Design}

Most of the LXeGRIT electronics is exposed to the environmental conditions
during the flight. The ambient pressure at float altitude is around 2~Torr, and
the temperature varies typically between $-20^\circ$~C during day time and
$-40^\circ$~C during night time, when the payload drops to lower
altitudes. During the ascent, the payload passes through even colder regions,
below $-60^\circ$~C. The heat produced by the circuits protects them from
getting too cold during this half an hour period. Once at float altitude, the
low pressure reduces the convection cooling by roughly a factor 20.  High power
circuits might overheat, if the produced heat is not efficiently transferred to
the aluminum structure of the payload. In addition, white panels shield the
gondola and its electronics from solar irradiation.  During the flight the
temperature of many critical parts is monitored by 16 temperature sensors.

The read-out processor together with its communication module incorporate
high power integrated circuits (IC). Providing an individual heat path for 
each IC would have been too difficult, therefore the 
processor box is hermetically sealed and kept under pressure. The 
CPU and one other circuit are responsible for most of the generated 
heat. They are, therefore, thermally grounded to the aluminum 
container. A miniature fan circulates the gas in the closed box, 
resulting in better heat transfer from the circuits on the 
board to the outside walls of the container. Hermetic connectors 
bring all the ports from the processor to the outside. A 
large aluminum heat sink on the outside of the container serves 
to increase the convection cooling in the thin atmosphere.

During the first flight of the new processor in 1999, the temperature in the
container stabilized at about \mbox{75\deg\ C}. Although the processor was
working well under these conditions, it was desirable to lower the operating
temperature. For the October 2000 flight of LXeGRIT, the container was therefore
filled with one atmosphere of helium. Due to the higher speed of the He
molecules, the heat transport to the outside walls is more efficient. This
resulted in a reduction of the operating temperature by almost 10 degrees.

The data storage disks also have to be mounted in a hermetic container, filled with
air under normal atmospheric pressure. This is not only for thermal
considerations, but also because they require an air cushion to separate the 
writing heads from the magnetic surface during operation. Heat conduction
through the mounting of the disks to their container ensured temperature
conditions well within specifications.

\section{Data Acquisition Software}

The implementation of a new read-out processor required the development of
a new DAQ software that would be able to take full advantage of the speed
of the processor and its various I/O interfaces. The software had to ensure
stable DAQ operation even under adverse conditions while aiming at maximum
data throughput from the digitizing hardware. Beyond efficiency, the design
was further required to be sufficiently flexible to adapt to the diverse
conditions during laboratory and flight operation, as well as to
allow the addition of new functionalities and upgrades, such as the
magnetometer read-out, added for the 2000 flight.

A key design choice was to rely on an embedded, multi-tasking, real-time
operating system (VxWorks from Wind River Systems): this provided a
complete, high-level framework of data structures and communication
mechanisms to accomplish task synchronization and I/O control, while
fulfilling the soft real-time requirements needed to saturate
hardware throughput (mostly quick reaction on the completion of direct
memory access $[$DMA$]$ transfers). Mission-critical robustness
requirements included that the operating system and the DAQ software can be
burnt on the local EE-PROM (electrically erasable programmable read-only
memory), allowing the system to boot and run even in case of disk
failure. In addition, the operating system also provided an efficient
interface during ground testing and software development.

\begin{figure*}
\centering
\includegraphics[bb=39 82 740 537,width=\textwidth,clip]{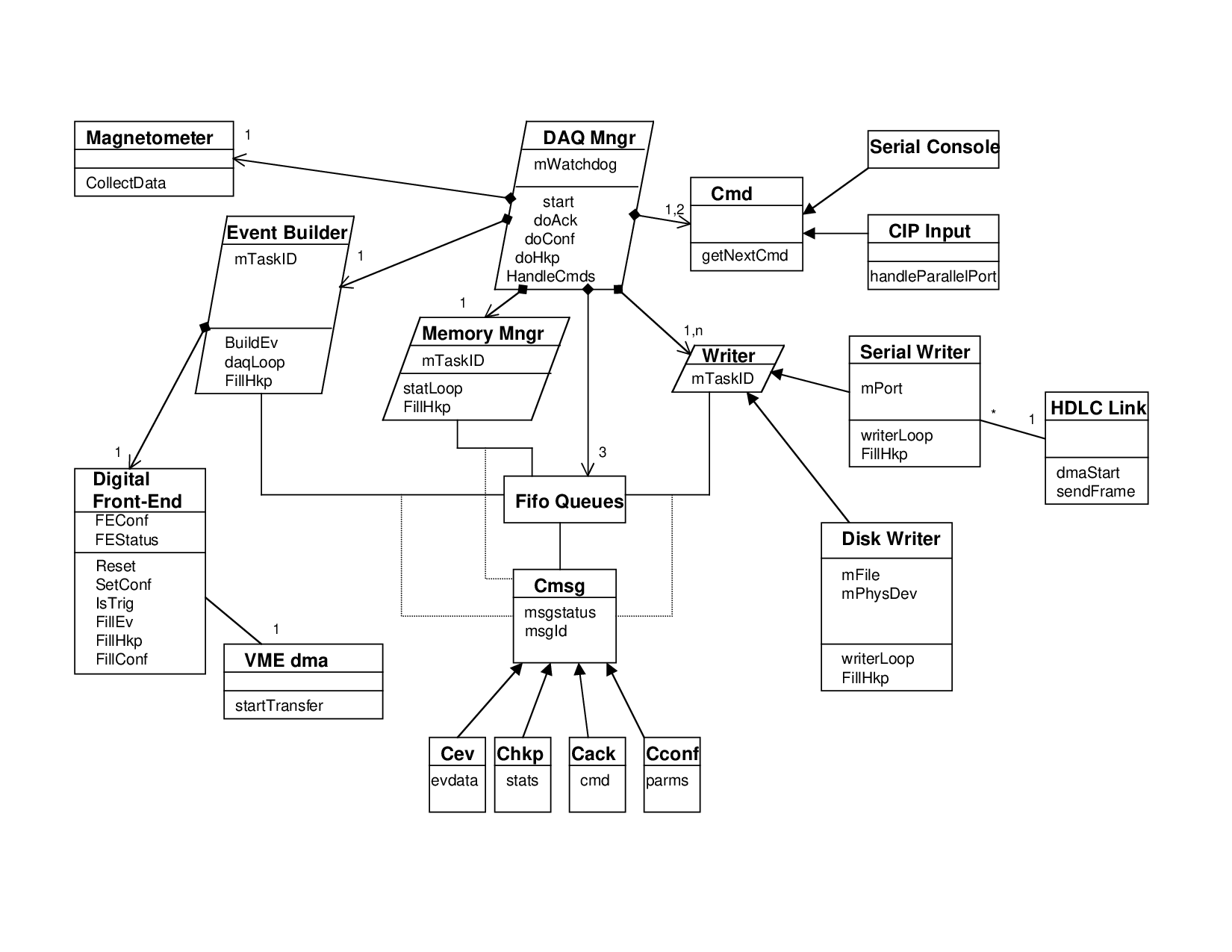}
\caption[]{\label{f:classdiag} 
Class diagram of the data acquisition software in flight 2000
configuration, adopting the UML (Unified Modeling Language)
notation \cite{BDouglass:98:UML}. Boxes represent classes, with three
compartments where appropriate: class name, attributes, and
methods. Parallelograms represent root classes, i.e., objects that are
mapped to separate threads. Solid lines without arrow (bidirectional) or
with an open arrow (unidirectional) indicate that one class uses another
class (association of classes), lines with filled diamonds indicate that
one class contains another class (aggregation of classes), and filled
arrows indicate that one class inherits another class
(specialization). Numbers indicate the number of related objects, where
``n'' means unknown at compile time and the asterisk means zero or
more. Cmsg is an associative class, defining the structure of the messages
sent via the FIFOs.
}
\end{figure*}

\begin{figure*}
\centering
\includegraphics[bb=53 111 746 549,width=\textwidth,clip]{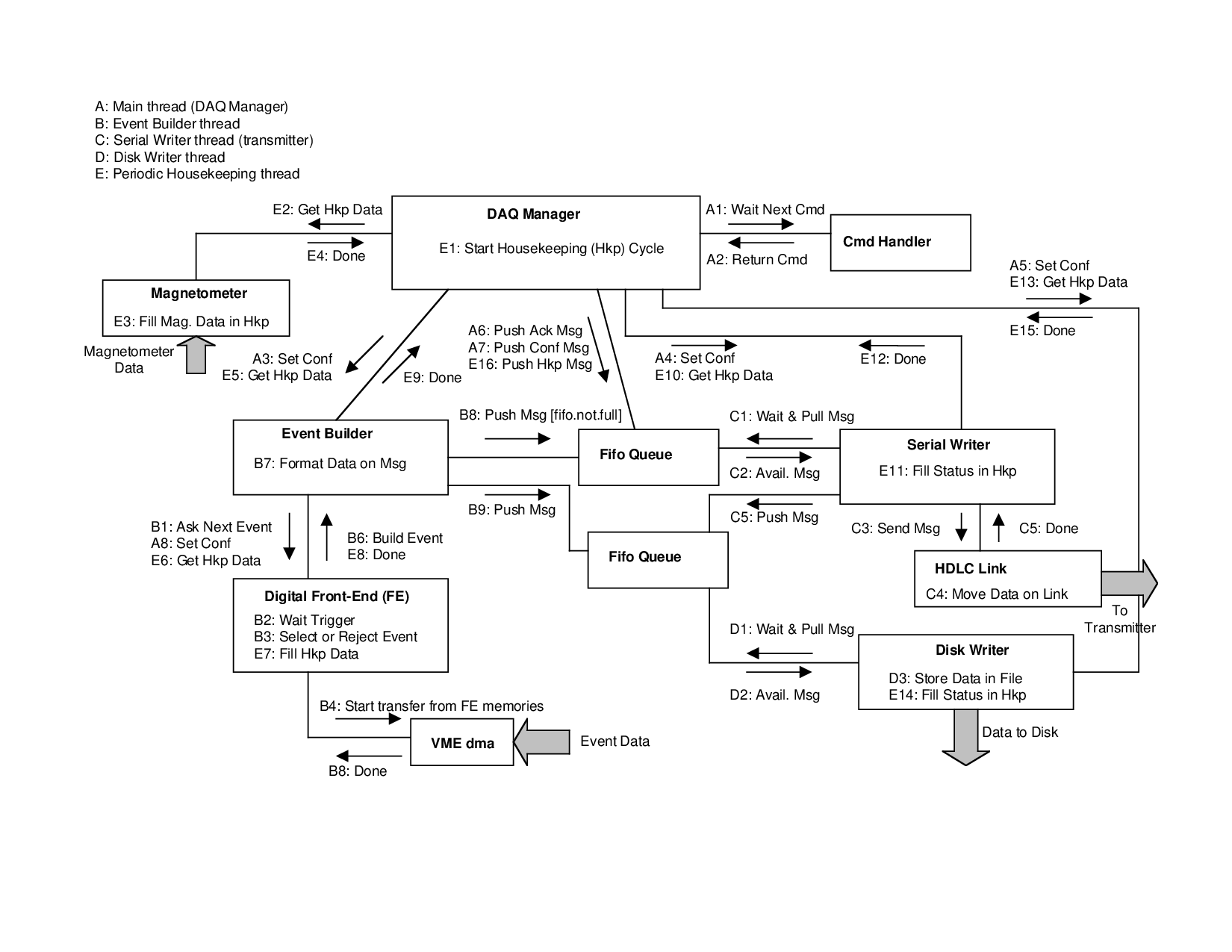}
\caption[]{\label{f:dataflow}
Diagram of the data acquisition software, describing the interactions among
objetcs and the data flow (collaboration diagram). Data are transferred mostly
through the FIFO queues, via messages (Msg) consisting of a header, specifying
its length and type, and a body, containing the data. For exchange of small
amounts of information between separate tasks, public interface routines of the
various classes are called to receive data (e.g., ``housekeeping'' data), to
receive commands, or to change configuration settings. Thick gray arrows
indicate the data generated in the front-end electronics and in the
magnetometer, which are eventually passed on to the local disks or to the
transmitters. A third FIFO, neglected in the figure for simplicity, is shared
among the Event Builder task and the Writer tasks and provides the Memory
Manager with the message pointers for memory deallocation.
}
\end{figure*}

An object-oriented software design kept strict independence among the
subsytems, which are connected to each other only through the sharing of
message queues and through well-defined interface routines (object
methods). This resulted in an easily reconfigurable system even at runtime,
where individual DAQ objects can be instantiated or deleted to adapt the
system to different conditions during the flight (e.g., turning disk
writing on/off) or during ground data taking (e.g., system control via
telemetry, serial console, or network).

Fig.~\ref{f:classdiag} depicts the class diagram of the DAQ software and
Fig.~\ref{f:dataflow} shows the data flows involved. The main subsystems are put
into individual threads (i.e., independent subprocesses of the main program), a
design which allows to optimize CPU usage and to balance the various tasks of
the read-out processor. Such tasks include science data acquisition, event
selection, collection of housekeeping data (see below), data transmission via
the serial links, disk writing, and providing a command interface to the user,
for DAQ control. Further optimization involved heavy use of the DMA engines
hosted on the processor board for all significant I/O operations, namely
read-out of event data, data transfer on the serial links, and disk
writing. This freed the CPU for event selection tasks and allowed the
various I/O operations to happen in parallel.

\subsection{DAQ Manager}
The main thread is the DAQ Manager. 
It configures the whole system and spawns all subprocesses.  It
receives control messages from the user via one of the commanding objects
and dispatches them to the appropriate system component.  The DAQ Manager
also spawns a housekeeping task every two seconds, a task which polls all
active objects to receive their status, collects rates from the
counters in the trigger electronics, and collects instrument attitude data
from the combined magnetometer / tiltmeter. After receiving a user
command, the DAQ Manager spawns a thread to build a command acknowledge
packet, and in case of changes in the data acquisition mode or on receiving
of a status inquiry, it also spawns a thread to build a detector
configuration packet. Those packets are inserted as messages in the FIFO
(first in first out) queue for subsequent forwarding both to the downlink
and to local hard disk storage.

\subsection{Event Builder}
The Event Builder thread aggregates all objects that interact with the data
acquisition hardware and specifies the sequence of operations following an event
trigger or a configuration message. It configures the hardware according to
default or user-supplied values, readies the waveform digitizers for event
collection, selects or rejects events after event triggers, and reads out signal
waveforms from the front-end memory banks. This thread is therefore the main
producer of data in the software system.  The event selection criteria are
user-configurable and aim at rejecting unwanted event topologies or empty/noisy
events, merely relying on a subset of the entire event information, such as the
number of wire hits or the energy deposited on the anodes. Accepted events are
delivered to the serial writers as messages through a shared FIFO queue, and, if
this queue is filled, through a second FIFO directly to the disk writer. If the
second FIFO is also filled, events are dispatched to a third FIFO queue, which
has been neglected in Fig.~\ref{f:dataflow} for simplicity. This queue is shared
among the Event Builder task and the Writer tasks, and provides the Memory
Manager with the message pointers for memory deallocation.  If even the third
queue is filled, the Event Builder suspends execution and waits until a buffer
in this queue becomes available.  This frees the CPU for the writer threads and
ensures, together with a higher priority given to the writer tasks, that the
Event Builder thread cannot stall the system, even if the trigger rate becomes
very large.

\subsection{Data Writers}
Two different classes of writer threads wait for and handle the
messages from the FIFO, either sending them via one of the two serial
links or saving them on one of the local hard disks. The messages can be of four
different types: a science event, housekeeping data, a command acknowledgement,
or a detector configuration packet. This architecture is very flexible since
several objects (and corresponding threads) can be working together on the same
FIFO, each dealing with sending data to a particular output. For instance, the
number of serial links can be reduced from two to one during runtime if one of
the transmitters fails. The serial writers transmit data in pieces of up to
2~kB, using an onboard chip that supplies HDLC (High-level Data Link Control)
framing, including a 32~bit cyclic redundancy check (CRC) word. This allows the
receiver on ground to check incoming data for completeness and accuracy. The
disk writer object not only stores data to disk but also takes care of the disk
management, switching to different disks as they fill up and switching power off
and on for idle and active disks.

\subsection{Memory Manager}
This independent thread provides memory management for a pool of buffers that
are preallocated at system startup, in order to avoid memory fragmentation. It
collects all the messages which went through the system before returning them to
the free buffer pool. Apart from computing statistics on data acquisition
performance and event selections, which are subsequently collected by the
housekeeping task, this thread was also meant to provide on-board data analysis
to reduce the amount of information to be downlinked or stored. This feature,
however, is presently disabled since the throughput of the writing channels in
the current system is larger than the front-end electronics throughput.

\section{System Performance}

The upgraded DAQ system has proven reliable during two flights in 1999 and
2000. For the 2000 flight, the development of an interface that allows block
transfers using the 
onboard DMA controller for reading of event data from the digitizer
electronics, as well as optimization of event selection criteria, has
accelerated the DAQ by a factor of about $2.5$ with respect to the 1999 flight
read-out. The data flow is now mainly limited by the interfacing of the
synchronous digitizer bus with the asynchronous VME bus,
which requires $\sim 1~\us$ per single byte access or $\sim 600$~ns using
block transfers. This sets a
limit on the total throughput of about 1.6~MB/s, restricting the event
building rate to 40 -- 50 events/s in ``full-image'' mode, in which the
complete digitized waveforms from all channels are to be read out,
amounting to about 30~kB per event. Standard data taking mode 
transfers only waveforms that crossed preset thresholds, plus the four anode
waveforms, and only for those events that fulfill a potential Compton
scattering topology, i.e., where the number of wire hits and therefore the
number of \g-ray interactions is within a preset range. In this mode, the
event build rate increases to 200 -- 400~Hz. The actual value
strongly depends on the selection parameters and on the light trigger
configuration, which also determine the average event size. For typical
settings, the average event size is 4.5 -- 6.5~kB.  

The throughput is sufficient to fill the two 500~kbps serial downlinks,
corresponding to about 2 full-image events/s per downlink, or about 10 -- 15
events/s in standard mode. The large bandwidth of disk writing, however, is
frequently not filled in standard mode, as event selections discard many
unwanted events in order to maximize the total number of triggers that can be
served by the system. During the October 2000 flight, the system was able to
handle about 300 -- 350~triggers/s out of a trigger rate of about 650~Hz, which
included charged particles since no plastic veto counters were used. About
20\% of the handled triggers were typically accepted as valid events, resulting
in a data throughput in the range of 0.4 -- 0.5~MB/s sent via the transmitters
and written to disk. In laboratory conditions with calibration sources, where
the ratio of accepted events is higher, this rate can become up to three times
larger.

\bibliographystyle{IEEEbib2}
\bibliography{mnemonic_short,myreferences}

\end{document}